\documentclass[11pt]{article}

\usepackage{latexsym,amsmath}

\DeclareFontFamily{OT1}{rsfs10}{}
\DeclareFontShape{OT1}{rsfs10}{m}{n}{ <-> rsfs10 }{}
\DeclareMathAlphabet{\mathscript}{OT1}{rsfs10}{m}{n}

\numberwithin{equation}{section}

\newcommand{\ns}{\normalsize}
\newcommand{\pt}{\partial}

\newcommand{\be}{\begin{equation}}
\newcommand{\ee}{\end{equation}}
\newcommand{\nn}{\nonumber}
\newcommand{\bea}{\begin{eqnarray}}
\newcommand{\eea}{\end{eqnarray}}
\newcommand{\tr}{\textrm{tr}}

\def\a{\alpha}
\def\b{\beta}
\def\g{\gamma}
\def\c{\chi}
\def\d{\delta}
\def\e{\epsilon}

\def\k{\kappa}

\def\m{\mu}
\def\n{\nu}

\def\p{\pi}

\def\r{\rho}
\def\s{\sigma}

\def\t{\tau}

\def\x{\xi}

\def\D{\Delta}

\def\cA{{\cal A}}
\def\cF{{\cal F}}
\def\cM{{\cal M}}

\begin{document}


\begin{titlepage}

\vspace{-2cm}

\title{
   \hfill{\ns UPR-805T, PUPT-1795\\}
   \hfill{\ns hep-th/9806022\\[.5cm]}
   {\LARGE Cosmological Solutions of Ho\v rava-Witten Theory}}
\author{
   Andr\'e Lukas$^1$
      \setcounter{footnote}{0}\thanks{Supported in part by Deutsche
          Forschungsgemeinschaft (DFG).}~~,
   Burt A.~Ovrut$^1$
      \setcounter{footnote}{3}\thanks{Supported in part by a Senior 
          Alexander von Humboldt Award.}~~,
   and Daniel Waldram$^2$\\[0.5cm]
   {\ns $^1$Department of Physics, University of Pennsylvania} \\
   {\ns Philadelphia, PA 19104-6396, USA}\\[0.3cm]
   {\ns $^2$Department of Physics, Joseph Henry Laboratories,}\\ 
   {\ns Princeton University, Princeton, NJ 08544, USA}}
\date{}

\maketitle

\begin{abstract}
We discuss cosmological solutions of Ho\v rava-Witten theory
describing the strongly coupled heterotic string. At energies below
the grand-unified scale, the effective theory is five- not
four-dimensional, where the additional coordinate parameterizes an
$S^1/Z_2$ orbifold. Furthermore, it admits no homogeneous
solutions. Rather, the static vacuum state, appropriate for a reduction to
four-dimensional $N=1$ supersymmetric models, is a BPS domain wall pair.
Relevant cosmological solutions are those associated with this BPS state. In
particular, such solutions must be inhomogeneous, depending on the
orbifold coordinate as well as on time. We present two examples of
this new type of cosmological solution, obtained by separation of
variables rather than by exchange of the time and radius coordinates
of a brane solution, as in previous work. The first example
represents the analog of a rolling radii solution with the radii
specifying the geometry of the domain wall pair. This is generalized in the
second example to include a nontrivial ``Ramond-Ramond'' scalar. 

\end{abstract}

\thispagestyle{empty}

\end{titlepage}


\section{Introduction}


Ho\v rava and Witten have shown that the strongly coupled $E_8\times E_8$ 
heterotic string can be identified as the eleven-dimensional limit of
M-theory compactified on an $S^1/Z_2$ orbifold with a set of $E_8$
gauge fields at each ten-dimensional orbifold fixed
plane~\cite{hw1,hw2}. Furthermore, Witten has demonstrated
that there exists a consistent compactification of this M-theory
limit on a ``deformed'' Calabi-Yau threefold, leading to a
supersymmetric $N=1$ theory in four dimensions~\cite{w}. Matching at
tree level to the phenomenological gravitational and grand-unified
(GUT)  couplings~\cite{w,bd}, one finds the orbifold must be larger
than the Calabi-Yau radius, by a factor of ten or so. Since the GUT
scale  (about $10^{16}$ GeV) is set by the size of the Calabi-Yau
threefold,  this implies that at energies below the unification scale
there is a regime where the universe appears five-dimensional.
This five-dimensional regime represents a new setting for
early universe cosmology, which has been traditionally studied in the
framework of the four-dimensional effective action. 

In a previous paper~\cite{losw}, the effective five-dimensional 
Ho\v rava-Witten theory was derived for the universal fields, which
are independent of the particular form of the Calabi--Yau manifold. It has a
number of interesting and unusual features. The theory lives in
a five-dimensional space which is a product of a smooth
four-dimensional manifold times the orbifold $S^1/Z_2$. 
As a result, it splits into a bulk $N=1$, $d=5$ supersymmetric theory
with the gravity supermultiplet and the universal hypermultiplet, and
two four-dimensional ``boundary'' theories which reside on the two
orbifold fixed hyperplanes. The additional fields of the
boundary theories are $N=1$, $d=4$ gauge multiplets and
chiral multiplets. The reduction from eleven-dimensional supergravity
to five dimensions requires the inclusion of non-zero values of the
four-form field strength in the internal Calabi-Yau directions. This
leads to a gauged version of five-dimensional supergravity with a
potential term that had not previously been constructed. In addition, 
the theory has boundary potentials for the projection of the bulk
scalar field onto the orbifold planes. 

These potentials lead to a particularly interesting
effect: the boundary sources mean that the theory has no solutions
homogeneous in the orbifold direction. In particular, flat space is
not a solution. Instead, the ``vacuum'' which leads to a supersymmetric
flat four-dimensional space is a three-brane domain wall solution. 
The three-brane couples to the bulk potential and is supported by the
sources provided by the two boundary potentials. More precisely, it is
a double domain wall solution with the two $3+1$-dimensional
worldvolumes each covering an orbifold plane and the orbifold itself
as the transverse coordinate. It is BPS, preserving half of the $d=5$
supersymmetries, and so is the appropriate background for a further
reduction to four-dimensional $N=1$ supergravity theories. In such a
reduction, four-dimensional space-time becomes identified with the
three-brane worldvolume.

Cosmologies in effective Ho\v rava-Witten theory should thus be both
{\em five-dimensional} and {\em inhomogeneous} in the extra
dimension. What should realistic models look like? In the ideal case,
one would have a situation in which the internal six-dimensional
Calabi-Yau space and the orbifold evolve in time for a short period
and then settle down to their ``phenomenological'' values while the
three non--compact dimensions continue to expand. Then, for late time,
when all physical scales are much larger than the orbifold size, the
theory is effectively four-dimensional and should, in the ``static''
limit, provide a realistic supergravity model of particle physics. As
we have argued above, such realistic supergravity models originate
from a reduction of the five-dimensional theory on its domain wall
vacuum state. Hence, in the ``static'' limit at late time, realistic
cosmological solutions should reduce to the domain wall or a perhaps a
modification thereof that incorporates breaking of the remaining
four-dimensional $N=1$ supersymmetry. Consequently, one is forced to
look for solutions which depend on the orbifold coordinate as well as
on time. The main goal of this paper is to present simple examples of
such cosmological solutions in five-dimensional heterotic M-theory to
illustrate some of the characteristic cosmological features of the
theory.

In earlier work~\cite{lowc1,lowc2}, we showed how a general class of
cosmological solutions, that is, time-dependent solutions of the
equations of motion that are homogeneous and isotropic in our physical
$d=3$ subspace, can be obtained in both superstring theories and
M-theory defined in spacetimes {\it without boundary}. Loosely
speaking, we showed that a cosmological solution could be obtained
from any p-brane or D-brane by inverting the roles of the time and
``radial'' spatial coordinate. This method will clearly continue to
work in Ho\v rava-Witten theory as long as one exchanges time with a
radial coordinate not aligned in the orbifold direction. An example of
this in eleven-dimensions, based on the solution of~\cite{llo}, has
been given in~\cite{benakli}. It can, however, not been applied to the
fundamental domain wall since its radial direction coincides with the
orbifold coordinate. This coordinate is bounded and cannot be turned
into time. Also, as argued above, exchanging radius and time in the
domain wall solution would not be desirable since it should be viewed
as the vacuum state and hence should not be modified in such a
way. Instead, the domain wall itself should be made time dependent
thereby leading to solutions that depend on both time and the orbifold
coordinate. As a result, we have to deal with coupled partial differential
equations, but, under certain constraints, these can by solved by separation
of variables, though the equations remain non-linear. Essentially, we are
allowing the moduli describing the geometry of the domain wall and the
excitations of other five-dimensional fields, to become
time-dependent. Technically, we will simply take the usual Ans\"atze
for the five-dimensional fields, but now allow the functions to depend
on {\it both} the time and radial coordinates. We will further demand
that these functions each factor into a purely time dependent piece
and a purely radial dependent piece. This is not, in general,
sufficient to separate the equations of motion. However, we will show
that subject to certain constraints separation of variables is
achieved. We can solve these separated equations and find new,
cosmologically relevant solutions. In this paper, we will restrict our
attention to two examples representing cosmological extensions of the
pure BPS three-brane. A more general class of solutions will be
presented elsewhere~\cite{llow}.

The first example is simply the domain wall itself with two of its three
moduli made time-dependent. We show that separation of variables occurs
in this case. It turns out that these moduli behave like
``rolling radii''~\cite{mueller} which constitute fundamental cosmological
solutions in weakly coupled string theory. Unlike those rolling radii
which represent scale factors of homogeneous, isotropic spaces, here they
measure the separation of the two walls of the three-brane and its
worldvolume size (which, at the same time, is the size of ``our''
three-dimensional universe). All in all, we therefore have a time-dependent
domain wall pair with its shape staying rigid but its size and separation
evolving like rolling radii. 

For the second example, we consider a similar setting as
for the first but, in addition, we allow a
nonvanishing ``Ramond-Ramond'' scalar. This terminology
is perhaps a little misleading, but relates to the fact that the
scalar would be a type II Ramond-Ramond field in the case where the
orbifold was replaced by a circle. This makes connection with
type II cosmologies with non-trivial Ramond-Ramond fields discussed
in~\cite{lowc1,lowc2}. Separation of variables occurs for a
specific time-independent form of this scalar. The
orbifold-dependent part then coincides with the domain wall with, however, the
addition of the Ramond-Ramond scalar. This non--vanishing value of the
scalar breaks supersymmetry even in the static limit.
We find that the time-dependent part of the equations fits into the
general scheme of M-theory cosmological solutions with form fields as
presented in ref.~\cite{lowc1,lowc2}. Applying the results of these
papers, the domain wall moduli are found to behave like rolling radii
asymptotically for early and late times. The evolution rates in these
asymptotic regions are different and the transitions between them can
be attributed to the nontrivial Ramond-Ramond scalar. 

Let us now summarize our conventions. We use coordinates $x^{\alpha}$
with indices $\alpha,\beta,\gamma,\cdots = 0,\cdots ,3,11$ to
parameterize the five-dimensional space $M_{5}$. Throughout this
paper, when we refer to the orbifold, we will work in the ``upstairs''
picture with the orbifold $S^1/Z_2$ in the $x^{11}$-direction. We
choose the range $x^{11}\in [-\pi\rho ,\pi\rho ]$ with the endpoints
being identified. The $Z_2$ orbifold symmetry acts as
$x^{11}\rightarrow -x^{11}$. Then there exist two four-dimensional
hyperplanes fixed under the $Z_2$ symmetry which we denote by
$M_{4}^{(i)}$, $i=1,2$. Locally, they are specified by the conditions
$x^{11}=0,\pi\rho$. The  indices $\mu,\nu,\rho,\cdots = 0,\cdots ,3$
are used for the four-dimensional space orthogonal to the
orbifold. Fields will be required to have a definite behaviour under
the $Z_2$ orbifold symmetry, so that a general field $\Phi$ is either
even or odd, with $\Phi (x^{11})=\pm\Phi (-x^{11})$.


\section{The five-dimensional effective action}


The five-dimensional effective action for Ho\v rava-Witten theory,
obtained from the eleven-dimensional theory by compactifying on a Calabi-Yau
three-fold, was derived in~\cite{losw} for the universal zero modes; that
is, the five-dimensional graviton supermultiplet and the breathing mode of
the Calabi-Yau space, along with its superpartners. These last fields form
a hypermultiplet in five dimensions. Furthermore, the theory contains
four-dimensional $N=1$ gauge multiplets and chiral gauge matter fields on
the orbifold planes. To keep the discussion as simple as possible, we do not
consider the latter. This simple framework suffices for the cosmological
solutions we will study in this paper. The general Lagrangian will be
presented elsewhere~\cite{pascos,losw1}.

In detail, we have the five-dimensional gravity supermultiplet
with the metric $g_{\a\b}$ and an Abelian gauge field $\cA_\a$ as the bosonic
fields. The bosonic fields in the universal hypermultiplet are the real
scalar field $V$ (the dilaton, measuring the volume of the internal
Calabi-Yau space), the three-form $C_{\a\b\g}$ and the complex Ramond-Ramond
scalar $\x$. Note that the three-form $C_{\alpha\beta\gamma}$ can be dualized
to a scalar field $\s$. Hence the hypermultiplet contains four real
scalar fields. As explained in the introduction, all bulk fields
should be even or odd under the $Z_2$ orbifold symmetry. One finds
that the fields $g_{\m\n}$, $g_{11,11}$, $\cA_{11}$, $\s$ must be even
whereas $g_{\m 11}$, $\cA_\m$, $\x$ must be odd. If one studies
cosmological solutions of the theory these transformation properties
are important as they restrict the set of allowed solutions to those
with the correct $Z_2$ symmetry. Now consider the boundary
theories. In the five-dimensional space $M_5$, the orbifold fixed
planes constitute the four-dimensional hypersurfaces $M_4^{(i)}$,
$i=1,2$. With the standard embedding in the reduction from eleven to five
dimensions, there will be an $E_6$ gauge field $A_\m^{(1)}$ and gauge
matter fields on the orbifold plane $M_4^{(1)}$. For simplicity, we
will set these gauge matter fields to zero in the following. On the
orbifold plane $M_4^{(2)}$ there is an $E_8$ gauge field
$A_\m^{(2)}$. 

The five-dimensional effective action of Ho\v rava-Witten theory is
then given by
\begin{equation}
 S_5 = S_{\rm bulk}+S_{\rm bound}\label{S5}
\end{equation}
where
\begin{align}
   S_{\rm bulk} =& -\frac{1}{2\k_5^2}\int_{M_5}\sqrt{-g}\left\{
        R + \frac{3}{2}\cF_{\a\b}\cF^{\a\b}
        + \frac{1}{\sqrt{2}}\e^{\a\b\g\d\e}\cA_\a\cF_{\b\g}\cF_{\d\e}
        + \frac{1}{2V^2}\partial_\a V\partial^\a V
     \right. \nn \\ & \quad \left. {}
        + \frac{1}{2V^2}
           \left[\pt_\a\s-i(\x\pt_\a\bar{\x}-\bar{\x}\pt_\a\x)
                 -2\a_0\e(x^{11})\cA_\a\right]
           \left[\pt^\a\s-i(\x\pt^\a\bar{\x}-\bar{\x}\pt^\a\x)
                 -2\a_0\e(x^{11})\cA^\a\right]
     \right. \nn \\ & \quad \left. {}
        + \frac{2}{V}\partial_\a\x\partial^\a\bar{\x}
        + \frac{1}{3V^2}\a_{0}^2\right\} 
   \\
   S_{\rm bound} =& \frac{\sqrt{2}}{\k_5^2}\int_{M_4^{(1)}}\sqrt{-g}\, 
           V^{-1}\a_{0}
         - \frac{\sqrt{2}}{\k_5^2}\int_{M_4^{(2)}}\sqrt{-g}\,
           V^{-1}\a_{0} \nn \\ & \quad
         - \frac{1}{16\p\a_{\rm GUT}}\sum_{i=1}^2
           \int_{M_4^{(i)}}\sqrt{-g}\, \left\{ 
              V \tr F_{\m\n}^{(i)}F^{(i)\m\n}  
              - \s \tr F_{\m\n}^{(i)}\tilde{F}^{(i)\m\n} \right\}
\; .\label{actparts}
\end{align}
where $\cF_{\a\b}=\partial_{\alpha}\cA_{\beta}-\partial_{\beta}\cA_{\alpha} 
$ and the $F^{(i)}_{\mu\nu}$ are the field strengths of the boundary gauge 
fields, while
$\tilde{F}^{(i)\m\n}=\frac{1}{2}\e^{\m\n\r\s}F_{\r\s}^{(i)}$.
Furthermore, $\k_5$ and $\a_{\rm GUT}$ are the five-dimensional 
Newton constant and the gauge coupling respectively. The constant
$\alpha_{0}$ in the above action can be computed for a given internal
Calabi-Yau space. Explicit formulae are presented in ref.~\cite{w,losw}.
In the above expression, we have dropped higher-derivative terms.
The sigma-model for the scalar fields is the well-known coset
$\cM_Q=SU(2,1)/SU(2)\times U(1)$ of the universal hypermultiplet. The
coupling of $\s$ to $\cA_\a$ implies that a $U(1)$ symmetry on
$\cM_Q$ has been gauged. This gauging also induces the
$\a_0$-dependent potential term in \eqref{actparts}. It has been
demonstrated~\cite{losw,losw1} that the above action is indeed the
bosonic part of a minimal $N=1$ gauged supergravity theory in five
dimensions coupled to chiral boundary theories.

The most striking features of this action from the viewpoint of cosmology
(and otherwise) are the bulk and boundary potentials for the dilaton $V$
in $S_{\rm bulk}$ and $S_{\rm bound}$. These potential terms are
proportional to the parameter $\alpha_{0}$ and their origin is directly
related to the nonzero internal four-form that had to be included in the
dimensional reduction from eleven dimensions. The boundary potentials lead
to sources in the Einstein equation and the equation of motion for $V$
and $\s$ that are proportional to $\d (x^{11})$ or $\d (x^{11}-\p\r
)$. Hence, as long as $V$ is finite (the internal Calabi-Yau space is
compact) purely time-dependent solutions of the theory do not exist
as they could never cancel these delta-function sources. One is
therefore led to always consider dependence on time and the orbifold
coordinate $x^{11}$. The presence of a bulk potential proportional to
$V^{-2}$ seems to indicate that the dilaton has a runaway behaviour
and the internal space decompactifies at late time. This picture,
however, is too naive in that it ignores the boundary potentials and
the $Z_2$ symmetries of the fields. In fact, as we will show, the
correct static domain wall vacuum of the theory depends on the
orbifold direction in a way so as to exactly cancel these
potentials. Consequently, it is important to note that for
cosmological solutions based on the domain wall the time-dependent
scale factors do not feel the potential terms.


\section{The domain-wall vacuum solution}


In this section, we would like to review the static ``vacuum'' solution 
of the five-dimensional Ho\v rava-Witten theory, as given
in~\cite{losw}. As argued in the introduction, this solution is the
basis for physically relevant 
cosmological solutions. It is clear from the 
five-dimensional action given in the previous section, that flat spacetime is 
not a solution of the equations of motion. It is precluded from being a 
solution by the potential terms, both in the bulk and on the boundaries.
If not flat space, what is the natural vacuum solution? To answer this,
notice that the theory~\eqref{S5} has all of the prerequisites necessary for
a three-brane solution to exist. Generally, in order to have a
$(D-2)$-brane in a $D$-dimensional theory, one needs to have a $(D-1)$-form
field or, equivalently, a cosmological constant. This is familiar from the
eight-brane~\cite{8brane} in the massive type IIA supergravity in ten
dimensions~\cite{romans}, and has been systematically studied for theories in
arbitrary dimension obtained by generalized (Scherk-Schwarz) dimensional
reduction~\cite{dom}. In our case, this cosmological term is provided by
the bulk potential term in the action~\eqref{S5}, precisely the term 
that disallowed flat space as a solution. From the
viewpoint of the bulk theory, we could have multi three-brane solutions with
an arbitrary number of parallel branes located at various places in the
$x^{11}$ direction. However, elementary brane solutions have
singularities at the location of the branes, needing to be supported by
source terms. The natural candidates for those source terms, in our case, are
the boundary actions. This restricts the possible solutions to those
representing a pair of parallel three-branes corresponding to the
orbifold planes. This pair of domain walls  can be viewed as the ``vacuum'' of
the five-dimensional theory, in the sense that it provides the appropriate
background for a reduction to the $d=4$, $N=1$ effective theory.

From the above discussion, it is clear that in order to find a three-brane
solution, we should start with the Ansatz
\bea
 ds_5^2 &=& a(y)^2dx^\m dx^\n\eta_{\m\n}+b(y)^2dy^2 \label{burt1} \\
 V &=& V(y)\nn
 \eea
where $a$ and $b$ are functions of $y=x^{11}$ and all other fields vanish.
The general solution for this Ansatz, satisfying the equations of motion
derived from action~\eqref{S5}, is given by
\bea
 a &=&a_0H^{1/2}\nn \\
 b &=& b_0H^2\qquad\qquad H=\frac{\sqrt{2}}{3}\a_{0}|y|+h_0 \label{burt2}\\
 V &=&b_0H^3 \nn
\eea
where $a_0$, $b_0$ and $h_0$ are constants. We note that the boundary
source terms have fixed the form of the harmonic function $H$ in the
above solution. Without specific information about the sources, the function
$H$ would generically be glued together from 
an arbitrary number of linear pieces with
slopes $\pm\sqrt{2}\a_{0}/3$. The edges of each piece would 
then indicate
the location of the source terms. The necessity of matching the boundary
sources at $y=0$ and $\p\r$, however, has forced us to consider only two such
linear pieces, namely $y\in [0,\p\r ]$ and $y\in [-\p\r ,0]$. These pieces are
glued together at $y=0$ and $\p\r$
(recall here that we have identified $\p\r$ and $-\p\r$).
To see this explicitly, let us consider one of the equations of motion; 
specifically, the equation derived from the variation of
$g_{\mu\nu}$. For the Ansatz in \eqref{burt1}, this is given by
\begin{equation}
 \frac{a''}{a}+\frac{{a'}^{2}}{a^{2}}-\frac{a'}{a}
  \frac{b'}{b}+\frac{1}{12}\frac{{V'}^{2}}{V^{2}}+
  \frac{\alpha_{0}^{2}}{18}\frac{b^{2}}{V^{2}} = \frac{\sqrt{2}\alpha_{0}}{3}
 \frac{b}{V}\left(\delta(y)-\delta(y-\pi\rho)\right)
\label{burt3}
\end{equation}
where the prime denotes differentiation with respect to $y$. The term 
involving the delta functions arises from the stress energy on the boundary 
planes.
Inserting the solution \eqref{burt2} in this equation, we
have 
\bea
\partial_y^2H &=& \frac{2\sqrt{2}}{3}\a_{0}(\d (y)-\d (y-\p\r )) 
\eea
which shows that the solution represents two parallel three-branes
located at the orbifold planes. Using the five-dimensional supersymmetry
transformations presented in ref.~\cite{losw}, one can check that this
solution indeed preserves four of the eight supersymmetries of the theory.

Let us discuss the meaning of this solution. As is apparent from the
Ansatz~\eqref{burt1}, it has $3+1$-dimensional Poincar\'e invariance
and, as just stated, it preserves four supercharges. Therefore, a
dimensional reduction to four dimensions on this solution leads to
an $N=1$ supergravity theory. In fact, this is just the ``physical''
four-dimensional effective theory of strongly coupled heterotic string
theory which is the starting point of low energy particle phenomenology.
This has been explicitly demonstrated in ref.~\cite{pascos,losw1}.
The two parallel three-branes of the solution, separated by the bulk,
are oriented in the four uncompactified space-time dimensions, and carry
the physical low-energy gauge and matter fields. Therefore, from the
low-energy point of view where the orbifold is not resolved, the 
three-brane worldvolume is identified with four-dimensional space-time.
In this sense the Universe lives on the worldvolume of a three-brane.
It is the purpose of the following sections to put this picture into
the context of cosmology; that is, to make it dynamical. Consequently,
we are looking for time dependent solutions based on the static domain
wall which we have just presented.


\section{The domain-wall cosmological solution}


In this section, we will present a cosmological solution related to the static
domain wall vacuum of the previous section. As discussed in
ref.~\cite{lowc1,lowc2}, a convenient way to find such a solution is to
use Ansatz \eqref{burt1} where the $y=x^{11}$ 
coordinate in the functions $a,b$ and $V$ is replaced by the time coordinate 
$\tau$. However, in Ho\v rava-Witten theory the boundary planes 
preclude this from being a solution of the equations of motion, since
it does not admit homogeneous solutions. To see this explicitly, let
us consider the $g_{00}$ equation of motion, where we replace $a(y)
\rightarrow \alpha(\tau), b(y) \rightarrow \beta(\tau)$ and $V
\rightarrow \gamma(\tau)$. We find that
\begin{equation}
 \frac{\dot{\alpha}^{2}}{\alpha^{2}}+\frac{\dot{\alpha}}{\alpha}
  \frac{\dot{\beta}}{\beta}-\frac{1}{12}\frac{\dot{\gamma}^{2}}{\gamma^{2}}
   -\frac{\alpha_{0}^{2}}{18}\frac{1}{\gamma^{2}}=
  -\frac{\sqrt{2}\alpha_{0}}{3}\frac{1}{\beta\gamma}\left(\delta(y)
  -\delta(y-\pi\rho)\right)\; ,
\label{burt4}
\end{equation}
where the dot denotes differentiation with respect to $\t$.
Again, the term containing the delta functions arises from the boundary planes.
It is clear that, because of the $y$-dependence introduced by the delta
functions, this equation has no globally defined solution. The structure of 
equation \eqref{burt4} suggests that a solution might be found if one were to 
let functions $a,b$ and $V$ depend on both $\tau$ and $y$ coordinates. This 
would be acceptable from the point of view of cosmology, since any such 
solution would be homogeneous and isotropic in the spatial coordinates $x^{m}$
where $m,n,r,\cdots = 1,2,3$. In fact, the previous Ansatz was too
homogeneous, being independent of the $y$ coordinate as well. Instead,
we are interested in solutions where the inhomogeneous vacuum domain
wall evolves in time. 

We now construct a cosmological 
solution where all functions depend on both $\tau$ and $y$. We start with the 
Ansatz
\bea
 ds_5^2 &=& -N(\tau,y)^{2}d\tau^{2}+a(\tau,y)^2dx^{m} dx^{n}\eta_{mn}+
            b(\tau,y)^2dy^2  \\
 V &=& V(\tau,y)\nn
 \label{burt5}
\eea
Note that we have introduced a separate function $N$ into the purely 
temporal part of the metric. 
This Ansatz leads to equations of motion that mix the $\tau$ and $y$
variables in a complicated non-linear way. In order to solve this system of 
equations, we will try to separate the two variables. That is, we let
\bea
 N(\tau,y)=n(\tau)a(y) \nn \\
 a(\tau,y)=\alpha(\tau)a(y) \nn\\
 b(\tau,y)=\beta(\tau)b(y) \label{burt6}\\
 V(\tau,y)=\gamma(\tau)V(y) \nn
\eea
There are two properties of this Ansatz that we wish to point out. 
The first is that for $n=\alpha=\beta=\gamma=1$ it becomes identical to
\eqref{burt1}. Secondly, we note that $n$ can be chosen to be any function 
by performing a redefinition of the $\tau$ variable. That is, we can 
think of $n$ as being subject to a gauge transformation. 
There is no a priori reason to believe that separation of variables will 
lead to a solution of the equations of motion derived from the
action~\eqref{S5}. However, as we now show, there is 
indeed such a solution. It is instructive to present one of the equations 
of motion. With the above Ansatz, the $g_{00}$ equation of motion is given
by~\footnote{From now on, we denote by $a$, $b$, $V$ the $y$-dependent
part of the Ansatz~\eqref{burt6}.}
\begin{multline}
\label{burt7}
 \frac{a^{2}}{b^{2}}\left(\frac{a''}{a}+\frac{{a'}^{2}}{a^{2}}-\frac{a'}{a}
  \frac{b'}{b}+\frac{1}{12}\frac{{V'}^{2}}{V^{2}}+\frac{\alpha_{0}^{2}}
   {18}\frac{b^{2}}{V^{2}}\frac{\beta^{2}}{\gamma^{2}}-\frac{\sqrt{2}}{3}
   \alpha_{0}\frac{b}{V}(\delta(y)-\delta(y-\pi\rho))\frac{\beta}{\gamma}
  \right) = \\
\frac{\beta^{2}}{n^{2}}\left(\frac{\dot{\alpha}^{2}}{\alpha^{2}}
 +\frac{\dot{\alpha}}{\alpha}\frac{\dot{\beta}}{\beta}-\frac{1}{12}
  \frac{\dot{\gamma}^{2}}{\gamma^{2}}\right)
\end{multline}
Note that if we set $n=\alpha=\beta=\gamma=1$ this equation becomes identical 
to \eqref{burt3}. Similarly, if we set $a=b=V=1$ and take the gauge $n=1$ 
this equation becomes the same as \eqref{burt4}. As is, the above 
equation does not separate. However, the obstruction to a separation of 
variables is the two terms proportional to $\alpha_{0}$. Note that both of 
these terms would be strictly functions of $y$ only if we demanded that
$\beta \propto \gamma$. Without loss of generality, one can take
\bea
\beta=\gamma\; .
\label{burt8}
\eea
We will, henceforth, assume this is the case. Note that this result is
already indicated by the structure of integration constants (moduli)
in the static domain wall solution~\eqref{burt2}. With this condition, the
left hand side of equation \eqref{burt7} is purely $y$ dependent, whereas the 
right hand side is purely $\tau$ dependent. Both sides must now equal the 
same constant which, for simplicity, we take to be zero. The equation 
obtained by setting the left hand side to zero is identical to the pure 
$y$ equation \eqref{burt3}. The equation for the pure $\tau$ dependent 
functions is
\bea
 \frac{\dot{\alpha}^{2}}{\alpha^{2}}
 +\frac{\dot{\alpha}}{\alpha}\frac{\dot{\beta}}{\beta}-\frac{1}{12}
  \frac{\dot{\gamma}^{2}}{\gamma^{2}}=0
\label{burt9}
\eea
Hence, separation of variables can be achieved for the $g_{00}$ equation by 
demanding that \eqref{burt8} is true. What is more remarkable is that, 
subject to the constraint that $\beta=\gamma$, all the equations of motion 
separate. The pure $y$ equations are identical to those of the previous 
section and, hence, the domain wall solution \eqref{burt2} remains valid
as the $y$--dependant part of the solution. 

The full set of 
$\tau$ equations is found to be
\bea
 \frac{\dot{\alpha}^{2}}{\alpha^{2}}
 +\frac{\dot{\alpha}}{\alpha}\frac{\dot{\beta}}{\beta}-\frac{1}{12}
  \frac{\dot{\gamma}^{2}}{\gamma^{2}}=0
\label{burt10}
\eea
\bea
 2\frac{\ddot{\alpha}}{\alpha}-2\frac{\dot{\alpha}}{\alpha}\frac{\dot{n}}{n}
  +\frac{\ddot{\beta}}{\beta}-\frac{\dot{\beta}}{\beta}\frac{\dot{n}}{n}
   +\frac{\dot{\alpha}^{2}}{\alpha^{2}}+2\frac{\dot{\alpha}}{\alpha}
   \frac{\dot{\beta}}{\beta}+\frac{1}{4}\frac{\dot{\gamma}^{2}}
  {\gamma^{2}}=0
\label{burt11}
\eea
\bea
 \frac{\ddot{\alpha}}{\alpha}-\frac{\dot{\alpha}}{\alpha}\frac{\dot{n}}{n}
  +\frac{\dot{\alpha}^{2}}{\alpha^{2}}+\frac{1}{12}\frac{\dot{\gamma}^{2}}
  {\gamma^{2}}=0
\label{burt12}
\eea
\bea
 \frac{\ddot{\gamma}}{\gamma}+3\frac{\dot{\a}\dot{\g}}{\a\g}+
  \frac{\dot{\b}\dot{\g}}{\b\g}-\frac{\dot{\g}^2}{\g^2}
  -\frac{\dot{n}\dot{\g}}{n\g}=0
\label{burt13}
\eea
In these equations we have displayed $\beta$ and $\gamma$ independently, 
for reasons to become apparent shortly. Of course, one must solve these 
equations subject to the condition that $\beta=\gamma$.
As a first attempt to solve these equations, it is most convenient to choose 
a gauge for which
\bea 
n=\mbox{const}
\label{burt14}
\eea
so that $\t$ becomes proportional to the comoving time $t$, since
$dt=n(\t )d\t$. In such a gauge, the equations simplify considerably and
we obtain the solution
\bea
 \a &=&A |t-t_0|^p \nn\\
 \b &=&\g = B |t-t_0|^q \label{burt15}
\eea
where
\begin{equation}
 p=\frac{3}{11}(1\mp\frac{4}{3\sqrt{3}})\; ,\qquad
 q=\frac{2}{11}(1\pm2\sqrt{3})
\label{burt16}
\end{equation}
and $A$, $B$ and $t_0$ are arbitrary constants. 
We have therefore found a cosmological solution, based on the separation
Ansatz~\eqref{burt6}, with the $y$-dependent
part being identical to the domain wall solution~\eqref{burt2} and the
scale factors $\a$, $\b$, $\g$ evolving according to the power
laws~\eqref{burt15}. This means that the shape of the domain wall pair stays
rigid while its size and the separation between the walls evolve in
time. Specifically, $\a$ measures
the size of the spatial domain wall worldvolume (the size of the
three-dimensional universe), while $\b$ specifies the separation of the
two walls (the size of the orbifold). Due to the separation constraint
$\g =\b$, the time evolution of the Calabi-Yau volume, specified by $\g$,
is always tracking the orbifold. From this point of view, we are
allowing two of the three moduli in~\eqref{burt2}, namely $a_0$ and
$b_0$, to become time-dependent. Since these moduli multiply the harmonic
function $H$, it is then easy to see why a solution by separation of
variables was appropriate. 

To understand the structure of the above solution, it is useful to
rewrite its time dependent part in a more systematic way using
the formalism developed in ref.~\cite{lowc1,lowc2}. First, let us define
new functions $\hat{\alpha}, \hat{\beta}$ and $\hat{\g}$ by
\begin{equation}
 \alpha=e^{\hat{\alpha}}, \qquad  \beta=e^{\hat{\beta}}, \qquad
 \gamma=e^{6\hat{\g}}
\label{burt17}
\end{equation}
and introduce the vector notation
\begin{equation}
 \vec{\alpha}= (\a^i) =\left(
            \begin{array}{c}
            \hat{\alpha} \\
            \hat{\beta} \\    
            \hat{\g} \\
            \end{array}
            \right)\; ,\qquad
 \vec{d}= (d_i) = \left(
            \begin{array}{c}
            3 \\
            1 \\    
            0 \\
            \end{array}
            \right)\; .
 \label{burt18}
\end{equation} 
Note that the vector $\vec{d}$ specifies the dimensions of the various
subspaces, where the entry $d_1=3$ is the spatial worldvolume dimensions,
$d_2=1$ is the orbifold dimension and we insert $0$ for the dilaton.
On the ``moduli space'' spanned by $\vec{\alpha}$ we introduce the
metric
\bea
 G_{ij}&=&2(d_{i}\delta_{ij}-d_{i}d_{j}) \nn \\
 G_{in}&=&G_{ni}=0 \label{andre5}\\
 G_{nn}&=&36\; ,\nn
\eea
which in our case explicitly reads
\begin{equation}
G=-12 \left(
      \begin{array}{ccc}
      1  & \frac{1}{2} & 0  \\
      \frac{1}{2} & 0  & 0  \\
      0  & 0           & -3 \\
      \end{array}
      \right)\; .
\label{burt20}
\end{equation}
Furthermore, we define $E$ by
\begin{equation}
 E=\frac{e^{\vec{d}\cdot\vec{\alpha}}}{n}=
   \frac{e^{3\hat{\alpha}+\hat{\beta}}}{n}\; .
\label{burt19}
\end{equation}
The equations of motion \eqref{burt10}-\eqref{burt13} can then be
rewritten as
\begin{equation}
 \frac{1}{2}E\dot{\vec{\alpha}}^{T}G\dot{\vec{\alpha}}=0\; ,\qquad
 \frac{d}{d\t}\left(EG\dot{\vec{\alpha}}\right)=0\; .
\label{burt22}
\end{equation}
It is straightforward to show that if we choose a gauge $n=$ const, these
two equations exactly reproduce the solution given in~\eqref{burt15}
and~\eqref{burt16}. The importance of this reformulation of the
equations of motion lies, however, in the fact that we now get solutions
more easily by exploiting the gauge choice for $n$. For example, let us
now choose the gauge
\begin{equation}
 n = e^{\vec{d}\cdot\vec{\a}}\; .
\end{equation}
Note that in this gauge $E=1$. The reader can verify that this gauge choice
greatly simplifies solving the equations. The result is that
\bea
 \hat{\a}&=&6\hat{\g}=C\t +k_1 \nn \\
 \hat{\b}&=&(6\pm 4\sqrt{3})C\t +k_2
\eea
where $C$, $k_1$ and $k_2$ are arbitrary constants. Of course this solution
is completely equivalent to the previous one, eq.~\eqref{burt15}, but written
in a different gauge. We will exploit this gauge freedom to effect in the
next section.

To discuss cosmological properties we define the Hubble parameters
\begin{equation}
 \vec{H}= \frac{d}{dt}\vec{\alpha}
\label{andre12}
\end{equation}
where $t$ is the comoving time. From \eqref{burt15} and \eqref{burt17}
we easily find
\begin{equation}
 \vec{H}=\frac{\vec{p}}{t-t_0}\; ,\qquad 
  \vec{p}= \left(
                \begin{array}{c}
                p \\
                q \\
                \frac{1}{6}q \\
                \end{array}
                \right)\; .
 \label{Hubble}
\end{equation}
Note that the powers $\vec{p}$ satisfy the constraints
\begin{equation}
 \vec{p}^{T}G\vec{p}=0 \qquad  \vec{d}\cdot \vec{p}=1\; .
\label{hello2}
\end{equation}
These relations are characteristic for rolling radii solutions~\cite{mueller}
which are fundamental cosmological solutions of weakly coupled heterotic
string theory. Comparison of the equations of motion~\eqref{burt22}
indeed shows that the scale factors $\vec{\a}$ behave like rolling
radii. The original rolling radii solutions describe freely evolving scale
factors of a product of homogeneous, isotropic spaces.
In our case, the scale factors also evolve freely (since the time-dependent
part of the equations of motion, obtained after separating variables, does
not contain a potential) but they describe the time evolution of the
domain wall. This also proves our earlier claim that the potential terms
in the five-dimensional action~\eqref{S5} do not directly influence the
time-dependence but are canceled by the static domain wall part of the
solution.

Let us now be more specific about the cosmological properties of our
solution. First note from eq.~\eqref{Hubble} that there exist two
different types of time ranges, namely $t<t_0$ and $t>t_0$. In the first
case, which we call the $(-)$ branch, the evolution starts at
$t\rightarrow -\infty$ and runs into a future curvature
singularity~\cite{lowc1,lowc2} at
$t=t_0$. In the second case, called the $(+)$ branch, we start out in
a past curvature singularity at $t=t_0$ and evolve toward
$t\rightarrow\infty$. In summary, we therefore have the branches
\begin{equation}
 t\in \left\{\begin{array}{cc}
             \mbox{[$-\infty,t_{0}$]} & (-)\;\mbox{branch} \\
             \mbox{[$t_{0},+\infty$]} & (+)\;\mbox{branch} \\
             \end{array}
             \right. \; .
\label{andre11}
\end{equation}
For both of these branches we have two options for the powers
$\vec{p}$, defined in eq.~\eqref{Hubble}, corresponding to the two
different signs in eq.~\eqref{burt16}. Numerically, we find
\begin{equation}
 \vec{p}_{\uparrow}\simeq \left(
                \begin{array}{c}
                +.06 \\
                +.81 \\
                +.14 \\
                \end{array}
                \right)\; ,\qquad
 \vec{p}_{\downarrow}\simeq \left(
                \begin{array}{c}
                +.48 \\
                -.45 \\
                -.08 \\
                \end{array}
                \right)
\label{andre16}
\end{equation}
for the upper and lower sign in~\eqref{burt16} respectively. We recall
that the three entries in these vectors specify the evolution powers for
the spatial worldvolume of the three-brane, the domain wall separation and
the Calabi-Yau volume. The expansion of the domain wall worldvolume has
so far been measured in terms of the five-dimensional Einstein frame
metric $g^{(5)}_{\m\n}$. This is also what the above numbers $p_1$ reflect.
Alternatively, one could measure this expansion
with the four-dimensional Einstein frame metric $g^{(4)}_{\m\n}$ so
that the curvature scalar on the worldvolume is canonically normalized.
From the relation
\begin{equation}
 g^{(4)}_{\m\n} = \left( g_{11,11}\right)^{1/2}g^{(5)}_{\m\n}
\end{equation}
we find that this modifies $p_1$ to
\begin{equation}
 \tilde{p}_1 = p_1+\frac{p_2}{2}\; .\label{conv}
\end{equation}
In the following, we will discuss both frames. We recall that the separation
condition $\b =\g$ implies that the internal Calabi-Yau space always
tracks the orbifold. In the discussion we can, therefore, concentrate on
the spatial worldvolume and the orbifold, corresponding to the first
and second entries in \eqref{andre16}. Let us first consider the
$(-)$ branch. In this branch $t\in [-\infty ,t_0]$ and, hence, $t-t_0$ is
always negative. It follows from eq.~\eqref{Hubble} that a subspace will
expand if its $\vec{p}$ component is negative and contract if it is
positive. For the first set of powers $\vec{p}_{\uparrow}$ in
eq.~\eqref{andre16} both the worldvolume and the orbifold contract in
the five-dimensional Einstein frame. The same conclusion holds in the
four-dimensional Einstein frame. For the second set, $\vec{p}_{\downarrow}$,
in both frames the worldvolume contracts while the orbifold expands.
Furthermore, since the Hubble parameter of the orbifold increases in time
the orbifold undergoes superinflation.

Now we turn to the $(+)$ branch. In this branch $t\in [t_0,\infty ]$ and,
hence, $t-t_0$ is always positive. Consequently, a subspace expands for
a positive component of $\vec{p}$ and contracts otherwise. In addition,
since the absolute values of all powers $\vec{p}$ are smaller than one,
an expansion is always subluminal. For the vector $\vec{p}_{\uparrow}$
the worldvolume and the orbifold expand in both frames. On the other
hand, the vector $\vec{p}_{\downarrow}$ describes an expanding worldvolume
and a contracting orbifold in both frames. This last solution perhaps
corresponds most closely to our notion of the early universe.


\section{Cosmological solutions with Ramond forms}


Thus far, we have looked for both static and cosmological solutions where the
form fields $\x, {\cal{A}}_{\a}$ and $\s$ have been set to zero.
As discussed in previous papers~\cite{lowc1,lowc2}, turning on one or
several such fields can drastically alter the solutions and their cosmological
properties. Hence, we would like to explore cosmological solutions with
such non-trivial fields. For clarity, in this paper we will restrict the
discussion to turning on the Ramond-Ramond scalar $\x$ only, postponing the
general discussion to another publication~\cite{llow}. 

The Ansatz we will use is the following. For the metric and dilaton field, we 
choose
\bea 
 ds_5^2 &=& -N(\tau,y)^{2}d\tau^{2}+a(\tau,y)^2dx^{m} dx^{n}\eta_{mn}+
            b(\tau,y)^2dy^2   \label{dan1}\\
 V &=& V(\tau,y)\; .\nn
\eea
For the $\x$ field, we assume that $\x=\x(\tau,y)$ and, hence, the field 
strength $F_{\a}=\partial_{\a}\xi$ is given by
\begin{equation}
 F_{0}=Y(\tau,y)\; ,\qquad  F_{5}=X(\tau,y)\; .
\label{dan2}
\end{equation}
All other components of $F_{\a}$ vanish. Note that since $\x$ is complex, both 
$X$ and $Y$ are complex. Once again, we will solve the equations of motion
by separation of variables. That is, we let 
\bea
 N(\tau,y)=n(\tau)N(y) \nn\\
 a(\tau,y)=\alpha(\tau)a(y) \nn\\
 b(\tau,y)=\beta(\tau)b(y) \label{dan3}\\
 V(\tau,y)=\gamma(\tau)V(y)\nn
\eea
and
\bea
 X(\tau,y)=\chi(\tau)X(y)\\
 Y(\tau,y)=\phi(\tau)Y(y)\; .
\label{dan4}
\eea
Note that, in addition to the $\x$ field, we have also allowed for
the possibility that $N(y)\neq a(y)$. Again, there is no a priori reason to 
believe that a solution can be found by separation of variables. However, as 
above, there is indeed such a solution, although the constraints required to 
separate variables are more subtle. It is instructive to present one of the 
equations of motion. With the above Ansatz, the $g_{00}$ equation of motion 
becomes~\footnote{In the following, $N$, $a$, $b$, $V$ denote the
$y$-dependent part of the Ansatz~\eqref{dan3}.}
\begin{multline}
\label{dan5}
 \frac{N^{2}}{b^{2}}\left(\frac{a''}{a}+\frac{{a'}^{2}}{a^{2}}-\frac{a'}{a}
  \frac{b'}{b}+\frac{1}{12}\frac{{V'}^{2}}{V^{2}}
  +\frac{\alpha_{0}^{2}}{18}\frac{b^{2}}{V^{2}}\frac{\beta^{2}}{\gamma^{2}}
  -\frac{\sqrt{2}\alpha_{0}}{3}
     \frac{b}{V}(\delta(y)-\delta(y-\pi\rho))\frac{\beta}{\gamma}
  \right) = \\
  \frac{\beta^{2}}{n^{2}}\left(\frac{\dot{\alpha}^{2}}{\alpha^{2}}
  +\frac{\dot{\alpha}}{\alpha}\frac{\dot{\beta}}{\beta}-\frac{1}{12}
  \frac{\dot{\gamma}^{2}}{\gamma^{2}}\right)
  -\frac{N^{2}}{3b^{2}}\frac{|X|^{2}}{V}\frac{|\c |^{2}}{\gamma}
  -\frac{\beta^{2}}{3n^{2}}\frac{|Y|^{2}}{V}\frac{|\phi |^{2}}{\gamma}
\end{multline}
Note that if we set $X=Y=0$ and $N=a$ this equation becomes identical to
\eqref{burt7}. We now see that there are two different types of obstructions 
to the separation of variables. The first type, which we encountered in the 
previous section, is in the two terms proportional to $\alpha_{0}$. 
Clearly, we can separate variables only if we demand that
\bea
\beta=\gamma
\label{dan6}
\eea
as we did previously. However, for non-vanishing $X$ and $Y$ this is not 
sufficient. The problem, of course, comes from the last two terms in \eqref
{dan5}. There are a number of options one could try in order to separate 
 variables in these terms. It is important to note that $X$ and $Y$ are not 
completely independent, but are related to each other by the integrability 
condition  $\partial_{\tau}X(\tau,y)=\partial_{y}Y(\tau,y)$. We find that, 
because of this condition, it is impossible to obtain a solution by 
separation of variables that has both $X(\tau,y)$ and $Y(\tau,y)$ 
non-vanishing. Now $X(\tau,y)$, but not $Y(\tau,y)$, can be made to vanish 
by taking $\x=\x(\tau)$; that is, $\x$ is a function of $\tau$ only. However
we can find no solution by separation of variables under this circumstance. 
Thus, we are finally led to the choice $\x=\x(y)$. In this case 
$Y(\tau,y)=0$ and we can, without lose of generality, choose
\bea
 \chi=1\; .
\label{dan7}
\eea
At this point, the only obstruction to separation of variables in equation
\eqref{dan5} is the next to last term, $N^2|X|^2/3b^2V\g$. Setting
$\gamma=\mbox{const}$ is too restrictive, so we must demand that 
\bea
  X = \frac{bV^{1/2}}{N}c_{0}e^{i\theta(y)}
\label{dan8}
\eea
where $c_{0}$ is a non-zero but otherwise arbitrary real constant and
$\theta(y)$ is an, as yet, undetermined phase.
Putting this condition into the $\x$ equation of motion
\bea
 \partial_{y}\left(\frac{a^3N}{bV}X\right)=0
\label{dan9}
\eea
we find that $\theta$ is a constant $\theta_0$ and $a\propto
V^{\frac{1}{6}}$ with arbitrary coefficient. Note that the last condition is 
consistent with the static vacuum solution \eqref{burt2}. Inserting this 
result into the $g_{05}$ equation of motion
\bea
 \frac{\dot{\alpha}}{\alpha}\left(\frac{a'}{a}-\frac{N'}{N}\right)=
 \frac{\dot{\beta}}{\beta}\left(\frac{a'}{a}-\frac{1}{6}\frac{V'}{V}\right)
\label{dan10}
\eea 
we learn that $N\propto a$ with arbitrary coefficient. Henceforth, we choose
$N=a$ which is consistent with the static vacuum solution \eqref{burt2}.
Inserting all of these results, the $g_{00}$ equation of motion now becomes
\begin{multline}
 \label{dan11}
 \frac{a^{2}}{b^{2}}\left(\frac{a''}{a}+\frac{{a'}^{2}}{a^{2}}-\frac{a'}{a}
  \frac{b'}{b}+\frac{1}{12}\frac{{V'}^{2}}{V^{2}}+\frac{\alpha_{0}^{2}}
   {18}\frac{b^{2}}{V^{2}}-\frac{\sqrt{2}}{3}
   \alpha_{0}\frac{b}{V}(\delta(y)-\delta(y-\pi\rho))\right) = \\
\frac{\beta^{2}}{n^{2}}\left(\frac{\dot{\alpha}^{2}}{\alpha^{2}}
 +\frac{\dot{\alpha}}{\alpha}\frac{\dot{\beta}}{\beta}-\frac{1}{12}
  \frac{\dot{\gamma}^{2}}{\gamma^{2}}\right)-\frac{c_{0}^{2}}{3}
  \frac{1}{\gamma}
\end{multline}
Note that the left hand side is of the same form as the static vacuum equation 
\eqref{burt3}. The effect of turning on the $\x$ background is to add a purely
$\tau$ dependent piece to the right hand side.
Putting these results into the remaining four equations of motion, we 
find that they too separate, with the left hand side being purely $y$ 
dependent and the right hand side purely $\tau$ dependent. Again, we 
find that in these equations the left hand sides are identical to those 
in the static vacuum equations and the effect of turning on $\x$ is to add
extra $\tau$ dependent terms to the right hand sides. In each equation,
both sides must now equal the 
same constant which, for simplicity, we take to be zero. The $y$ equations 
for $a$, $b$ and $V$ thus obtained by setting the left hand side to zero
are identical to the static vacuum equations. Hence, we have shown that
\bea
 N=a &=&a_0H^{1/2}\nn \\
 b &=& b_0H^2\qquad\qquad H=\frac{\sqrt{2}}{3}\a_{0}|y|+h_0 \label{dan12}\\
 V &=&b_0H^3 \nn \\
 X &=& x_{0}H^{3} \nn
\eea
where $x_{0}=c_{0}e^{i\theta_0}a_{0}^{-1}b_{0}^{3/2}$ is an arbitrary
constant. 

The $\tau$ equations obtained by setting the right hand side to zero are the 
following.
\bea
 \frac{\dot{\alpha}^{2}}{\alpha^{2}}
 +\frac{\dot{\alpha}}{\alpha}\frac{\dot{\beta}}{\beta}-\frac{1}{12}
  \frac{\dot{\gamma}^{2}}{\gamma^{2}}-\frac{c_{0}^{2}}{3}
  \frac{n^{2}}{\beta^{2}\gamma}=0
\label{dan13}
\eea
\bea
 2\frac{\ddot{\alpha}}{\alpha}-2\frac{\dot{\alpha}}{\alpha}\frac{\dot{n}}{n}
  +\frac{\ddot{\beta}}{\beta}-\frac{\dot{\beta}}{\beta}\frac{\dot{n}}{n}
   +\frac{\dot{\alpha}^{2}}{\alpha^{2}}+2\frac{\dot{\alpha}}{\alpha}
   \frac{\dot{\beta}}{\beta}+\frac{1}{4}\frac{\dot{\gamma}^{2}}
  {\gamma^{2}}-c_{0}^{2}\frac{n^{2}}{\beta^{2}\gamma} =0
\label{dan14}
\eea
\bea
 \frac{\ddot{\alpha}}{\alpha}-\frac{\dot{\alpha}}{\alpha}\frac{\dot{n}}{n}
  +\frac{\dot{\alpha}^{2}}{\alpha^{2}}+\frac{1}{12}\frac{\dot{\gamma}^{2}}
  {\gamma^{2}}+\frac{c_{0}^{2}}{3}\frac{n^{2}}{\beta^{2}\gamma} =0
\label{dan15}
\eea
\bea
 \frac{\ddot{\gamma}}{\gamma}+3\frac{\dot{\a}\dot{\g}}{\a\g}+
 \frac{\dot{\b}\dot{\g}}{\b\g}-\frac{\dot{\g}^2}{\g^2}
  -\frac{\dot{n}\dot{\g}}{n\g}
  -2c_{0}^{2}\frac{n^{2}}{\beta^{2}\dot{\gamma}} =0
\label{dan16}
\eea
In these equations we have, once again, displayed $\beta$ and $\gamma$ 
independently, although they should be solved subject to the 
condition $\beta=\gamma$. Note that the above 
equations are similar to the $\tau$ equations in the previous section, 
but each now has an additional term proportional to $c_{0}^{2}$. These 
extra terms considerably complicate finding a solution of the $\tau$ 
equations. Here, however, is where the formalism introduced in the 
previous section becomes important. Defining $\hat{\alpha},\hat{\beta}$ 
and $\hat{\g}$ as in \eqref{burt17}, and $\vec{\alpha}, E$ and $G$
as in \eqref{burt18}, \eqref{burt19} and \eqref{burt20} respectively,
the equations \eqref{dan13}-\eqref{dan16} can be written in the form
\begin{equation}
 \frac{1}{2}E\dot{\vec{\alpha}}^{T}G\dot{\vec{\alpha}}+E^{-1}U=0\; ,\qquad
 \frac{d}{d\t}\left(EG\dot{\vec{\alpha}}\right)+E^{-1}\frac{\partial U}
 {\partial\vec{\alpha}}=0
\label{dan18}
\end{equation}
where the potential $U$ is defined as
\begin{equation}
 U=2c_{0}^{2}e^{\vec{q}\cdot\vec{\alpha}}
\label{dan19}
\end{equation}
with
\begin{equation}
 \vec{q}= \left(
                \begin{array}{c}
                6 \\
                0 \\
                -6 \\
                \end{array}
                \right)\; .
\label{dan20}
\end{equation}
We can now exploit the gauge freedom of $n$ to simplify these equations.
Choose the gauge
\begin{equation}
 n=e^{(\vec{d}-\vec{q})\cdot\vec{\alpha}}
\label{dan21}
\end{equation}
where $\vec{d}$ is defined in \eqref{burt18}. Then
$E$ becomes proportional to the potential $U$ so that the potential
terms in \eqref{dan18} turn into constants. Thanks to this simplification,
the equations of motion can be integrated which leads to the
general solution~\cite{lowc1,lowc2}
\bea
 \vec{\alpha}=\vec{c}\,\ln|\tau_{1}-\tau|+\vec{w}\,\ln\left(\frac{s\tau}
 {\tau_{1}-\tau}\right) +\vec{k}
\label{dan24}
\eea
where $\tau_{1}$ is an arbitrary parameter which we take, without loss of 
generality, to be positive and
\begin{equation}
 \vec{c}=2\frac{G^{-1}\vec{q}}{<\vec{q},\vec{q}>}\; ,\qquad
 s=\mbox{sign}(<\vec{q},\vec{q}>)\; .
\label{dan25}
\end{equation}
The scalar product is defined as
$<\vec{q},\vec{q}>=\vec{q}^{T}G^{-1}\vec{q}$. The vectors $\vec{w}$ and 
$\vec{k}$ are integration constants subject to the constraints
\bea
 \vec{q}\cdot \vec{w}&=&1 \nn \\
 \vec{w}^{T}G\vec{w}&=&0 \label{dan25p}\\
 \vec{q}\cdot \vec{k}&=& \ln\left(c_{0}^{2}|<\vec{q},\vec{q}>|\right) \nn
\eea
This solution is quite general in that it describes an arbitrary number
of scale factors with equations of motion given by~\eqref{dan18}. Let us now
specify to our example. For $G$ and $\vec{q}$ as given in eq.~\eqref{burt20}
and \eqref{dan20} we find that
\begin{equation}
 <\vec{q},\vec{q}>=1
\label{dan26}
\end{equation}
hence $s=1$, and
\begin{equation}
 \vec{c}= \left(
                \begin{array}{c}
                0 \\
                -2 \\
                -\frac{1}{3} \\
                \end{array}
                \right)\; .
\label{dan27}
\end{equation}
Recall that we must, in addition, demand that $\beta=\gamma$. Note that the 
last two components of $\vec{c}$ are consistent with this equality.
We can also solve the constraints~\eqref{dan25p} subject to the
condition $\beta=\gamma$. The result is
\begin{equation}
 \vec{w}=     \left(
                \begin{array}{c}
                w_{3}+\frac{1}{6} \\
                6w_{3} \\
                w_{3} \\
                \end{array}
                \right)\; ,\qquad
 \vec{k}=   \left(
                \begin{array}{c}
                k_{3}+\frac{1}{6}\ln c_{0}^{2} \\
                6k_{3} \\
                k_{3} \\
                \end{array}
                \right)  
\label{dan29}
\end{equation}
where
\begin{equation}
 w_{3}=-\frac{1}{6}\pm\frac{\sqrt{3}}{12}
\end{equation}
and $k_{3}$ is arbitrary. We conclude that in the gauge specified by 
\eqref{dan21}, the solution is given by
\bea
 \hat{\alpha}=(w_{3}+\frac{1}{6})\ln\left(\frac{\tau}{\tau_{1}-\tau}\right)
 +k_{3}+\frac{1}{6}\ln c_{0}^{2} \nn\\
 \hat{\beta}= -2\ln |\tau_{1}-\tau| +6w_{3}\ln\left(\frac{\tau}
 {\tau_{1}-\tau}\right)+6k_{3} \label{dan30}\\
 \hat{\gamma}=-\frac{1}{3}\ln |\tau_{1}-\tau|+w_{3}\ln\left(\frac{\tau}
 {\tau_{1}-\tau}\right)+k_{3}\nn
\eea
with $w_{3}$ as above. As a consequence of $s=1$, the range for $\t$
is restricted to
\begin{equation}
 0<\tau<\tau_{1}
\label{dan34}
\end{equation}
in this solution. Let us now summarize our result. We have found a
cosmological solution with a nontrivial Ramond-Ramond scalar $\x$
starting with the separation Ansatz~\eqref{dan3}. To achieve separation
of variables we had to demand that $\b =\g$, as previously, and
that the Ramond-Ramond scalar depends on the orbifold coordinate but not on
time. Then the orbifold dependent part of the solution is given by
eq.~\eqref{dan12} and is identical to the static domain wall solution
with the addition of the Ramond-Ramond scalar. The time dependent
part, in the gauge~\eqref{dan21}, is specified by eq.~\eqref{dan30}.
Furthermore, we have found that the time-dependent part of the equations
of motion can be cast in a form familiar from cosmological solutions
studied previously~\cite{lowc1,lowc2}.
Those solutions describe the evolution for
scale factors of homogeneous, isotropic subspaces in the presence of
antisymmetric tensor fields and are, therefore, natural generalizations of
the rolling radii solutions. Each antisymmetric tensor field introduces
an exponential type potential similar to the one in eq.~\eqref{dan19}. For
the case with only one nontrivial form field, the general solution could
be found and is given by eq.~\eqref{dan24}. We have, therefore, constructed
a strong coupling version of these generalized rolling radii solutions
with a one-form field strength, where the radii now specify the domain wall 
geometry rather that the size of maximally symmetric subspaces. We stress
that the potential $U$ in the time-dependent equations of motion does
not originate from the potentials in the action~\eqref{S5} but from
the nontrivial Ramond-Ramond scalar. The potentials in the action are
canceled by the static domain wall part of the solution, as in the
previous example. 

From the similarity to the known generalized rolling
radii solutions, we can also directly infer some of the basic
cosmological properties of our solution, using the results of
ref.~\cite{lowc1,lowc2}. We expect the integration constants to split
into two disjunct sets which lead to solutions in the $(-)$ branch,
comoving time range $t\in [-\infty ,t_0]$, and the $(+)$ branch,
comoving time range $t\in [t_0,\infty ]$, respectively. The $(-)$ branch
ends in a future curvature singularity and the $(+)$ branch starts
in a past curvature singularity. In both branches the solutions behave
like rolling radii solutions asymptotically; that is, at
$t\rightarrow -\infty ,t_0$ in the $(-)$ branch and at 
$t\rightarrow t_0,\infty$ in the $(+)$ branch. The two asymptotic
regions in both branches have different expansion properties in general
and the transition between them can be attributed to the nontrivial
form field.

Let us now analyze this in more detail for our solution, following the
method presented in ref.~\cite{lowc1,lowc2}. First we should
express our solution in terms of the comoving time $t$ by integrating
$dt=n(\t )d\t$. The gauge parameter $n(\t )$ is explicitly given by
\begin{equation}
 n=e^{(\vec{d}-\vec{q})\cdot\vec{k}}|\tau_{1}-\tau|^{-x+\Delta-1}|\tau|^{x-1}
\label{dan33}
\end{equation}
where
\begin{equation}
 x=\vec{d}\cdot\vec{w}\; , \qquad\Delta=\vec{d}\cdot\vec{c}\; .
\label{dan31}
\end{equation}
Given this expression, the integration cannot easily be performed in general
except in the asymptotic regions $\t\rightarrow 0,\t_1$. These regions
will turn out to be precisely the asymptotic rolling-radii limits. Therefore,
for our purpose, it suffices to concentrate on those regions.
Eq.~\eqref{dan33} shows that the resulting range for the comoving time
depends on the magnitude of $\Delta$ and $x$ (note that $\Delta$ is a fixed
number, for a given model, whereas $x$ depends on the integration constants).
It turns out that for all values of the integration constants we have either
$x<\Delta$ or $x>0>\Delta$. This splits the space of integration
constant into two disjunct sets corresponding to the $(-)$ and the
$(+)$ branch as explained before. More precisely, we have the mapping
\begin{equation}
 \t\rightarrow t\in\left\{\begin{array}{clll}
       \left[ -\infty ,t_0\right]&{\rm for}\; x<\D<0\; ,&(-)\;{\rm branch} \\
       \left[ t_0,+\infty\right]&{\rm for}\; x>0>\D\; ,&(+)\;{\rm branch}
       \end{array}\right.
\label{mapping}
\end{equation}
where $t_0$ is a finite arbitrary time (which can be different for the
two branches). We recall that the range of $\t$ is $0<\t <\t_1$. The
above result can be easily read off from the expression~\eqref{dan33}
for the gauge parameter. Performing the integration in the asymptotic
region we can express $\t$ in terms of the comoving time and find
the Hubble parameters, defined by eq.~\eqref{andre12}, and the
powers $\vec{p}$. Generally, we have
\begin{equation}
 \vec{p} = \left\{\begin{array}{cll} \frac{\vec{w}}{x}&{\rm at}&\t\simeq 0 \\
                       \frac{\vec{w} -\vec{c}}{x-\D}&{\rm at}&\t\simeq\t_1
       \end{array}\right. \; . \label{p_expr}
\end{equation}
Note that, from the mapping~\eqref{mapping}, the expression at $\t\simeq 0$
describes the evolution powers at $t\rightarrow -\infty$ in the $(-)$
branch and at $t\simeq t_0$ in the $(+)$ branch; that is, the evolution
powers in the early asymptotic region. Correspondingly, the expression
for $\t\simeq\t_1$ applies to the late asymptotic regions; that is,
to $t\simeq t_0$ in the $(-)$ branch and to $t\rightarrow\infty$ in the
$(+)$ branch. As before, these powers satisfy the rolling
radii constraints~\eqref{hello2}.

Let us now insert the explicit expression for $\vec{d}$, $\vec{w}$ and
$\vec{c}$, eqs.~\eqref{burt18},\eqref{dan29} and \eqref{dan27}, that specify
our example into those formulae. First, from eq.~\eqref{dan31} we find that
\begin{equation}
 x=-1\pm3\frac{\sqrt{3}}{4}\; ,\qquad
 \Delta=-2\; .
\label{dan32}
\end{equation}
Note that the space of integration constants just consists of two points
in our case, represented by the two signs in the expression for $x$ above.
Clearly, from the criterion~\eqref{mapping} the upper sign leads to
a solution in the $(+)$ branch and the lower sign to a solution in the
$(-)$ branch. In each branch we therefore have a uniquely determined
solution. Using eq.~\eqref{p_expr} we can calculate the asymptotic
evolution powers in the $(-)$ branch
\begin{equation}
 \vec{p}_{-,t\rightarrow -\infty}=\left(
                \begin{array}{c}
                +.06 \\
                +.81 \\
                +.13 \\
                \end{array}
                \right)\; ,\qquad
 \vec{p}_{-,t\rightarrow t_0}=\left(
                \begin{array}{c}
                +.48 \\
                -.45 \\
                -.08 \\
                \end{array}
                \right)\; .
\label{kelly14}
\end{equation}
Correspondingly, for the $(+)$ branch we have
\begin{equation}
 \vec{p}_{+,t\rightarrow t_0}=\left(
                \begin{array}{c}
                +.48 \\
                -.45 \\
                -.08 \\
                \end{array}
                \right)\; ,\qquad
 \vec{p}_{+,t\rightarrow\infty}=\left(
                \begin{array}{c}
                +.06 \\
                +.81 \\
                +.13 \\
                \end{array}
                \right)\; .
\end{equation}
Note that these vectors are in fact the same as in the $(-)$
branch, with the time order being reversed. This happens because they
are three conditions on the powers $\vec{p}$ that hold in both branches,
namely the two rolling radii constraints~\eqref{hello2} and the separation
constraint $\b =\g$, eq.~\eqref{dan6}, which implies that $p_3=6p_2$.
Since two of these conditions are linear and one is quadratic, we expect
at most two different solutions for $\vec{p}$.
As in the previous solution, the time variation of the Calabi-Yau volume
(third entry) is tracking the orbifold variation (second entry) as a
consequence of the separation condition and, hence, needs not to be
discussed separately. The first entry gives the evolution power for the
spatial worldvolume in the five-dimensional Einstein frame. For a conversion
to the four-dimensional Einstein frame one should again apply
eq.~\eqref{conv}. It is clear from the above numbers, however, that this
conversion does not change the qualitative behaviour of the worldvolume
evolution in any of the cases. Having said this, let us first discuss
the $(-)$ branch. At $t\rightarrow -\infty$ the powers are
positive and, hence, the worldvolume and the orbifold are contracting.
The solution then undergoes the transition induced by the Ramond-Ramond
scalar. Then at $t\simeq t_0$ the worldvolume is still contracting while
the orbifold has turned into superinflating expansion.
In the $(+)$ branch we start out with a subluminally expanding worldvolume
and a contracting orbifold at $t\simeq t_0$. After the transition both
subspaces have turned into subluminal expansion.

\section{Conclusion}

In this paper we have presented the first examples of cosmological
solutions in five-dimensional Ho\v rava-Witten theory. They are
physically relevant in that they are related to the exact BPS three--brane
pair in five dimensions, whose $D=4$ worldvolume theory exhibits $N=1$
supersymmetry. A wider class of such cosmological solutions can be
obtained and will be presented elsewhere~\cite{llow}. We expect solutions
of this type to provide the fundamental scaffolding for theories of the
early universe derived from Ho\v rava-Witten theory, but they are clearly
not sufficient as they stand. The most notable deficiency is the fact
that they are vacuum solutions, devoid of any matter, radiation or
potential stress-energy. Inclusion of such stress-energy is essential
to understand the behaviour of early universe cosmology. A study of its
effect on the cosmology of Ho\v rava-Witten theory is presently 
underway~\cite{lownext}.

\vspace{0.4cm}

{\bf Acknowledgments} 
A.~L.~is supported in part by a fellowship from Deutsche
Forschungsgemeinschaft DFG). A.~L.~and B.~A.~O.~are supported in part by 
DOE under contract No. DE-AC02-76-ER-03071. D.~W.~is supported in part by
DOE under contract No. DE-FG02-91ER40671. 



\end{document}